\documentclass[twocolumn]{article}
\usepackage[a4paper, total={7in, 9in}]{geometry}

\usepackage{authblk}
\usepackage{graphicx}
\usepackage{amsmath, amssymb}
\usepackage{multirow}
\usepackage{xcolor}

\usepackage{tabularx}
\newcolumntype{s}{>{\hsize=.3\hsize}X}
\usepackage{booktabs}

\usepackage{tikz}
\usetikzlibrary{decorations.markings}

\usepackage{caption}
\usepackage{subcaption}

\usepackage[breaklinks]{hyperref}

\usepackage[capitalise]{cleveref}
\crefname{app}{Appendix}{Appendices}

\usepackage[sorting=none, style=numeric-comp, maxnames=99]{biblatex}
\title{Reinforcement Learning for Adaptive Composition of Quantum Circuit Optimisation Passes}

\author[1]{Daniel Mills}
\author[1]{Ifan Williams\thanks{Corresponding author: ifan.williams@quantinuum.com}}
\author[2]{Jacob Swain}
\author[2]{Gabriel Matos}
\author[2]{Enrico Rinaldi}
\author[2]{Alexander Koziell-Pipe}
\affil[1]{Quantinuum, Terrington House, 13–15 Hills Road, Cambridge CB2 1NL, United Kingdom}
\affil[2]{Quantinuum, Partnership House, Carlisle Place, London SW1P 1BX, United Kingdom}

\date{}

\begin{document}

\maketitle

\begin{abstract}
    Many quantum software development kits provide a suite of circuit optimisation passes. These passes have been highly optimised and tested in isolation. However, the order in which they are applied is left to the user, or else defined in general-purpose default pass sequences. While general-purpose sequences miss opportunities for optimisation which are particular to individual circuits, designing pass sequences bespoke to particular circuits requires exceptional knowledge about quantum circuit design and optimisation. Here we propose and demonstrate training a reinforcement learning agent to compose optimisation-pass sequences. In particular the agent's action space consists of passes for two-qubit gate count reduction used in default PyTKET pass sequences.
    For the circuits in our diverse test set, the (mean, median) fraction of two-qubit gates removed by the agent is $(57.7\%, \ 56.7 \%)$, compared to $(41.8 \%, \ 50.0 \%)$ for the next best default pass sequence.
    In doing so our model demonstrates the ability to forgo reward to improve the final two-qubit gate count. Our approach clearly separates the problem of designing circuit optimisation passes from the problem of using them optimally in practice. We expect this separation of concerns to facilitate innovations in both areas.
\end{abstract}

\section{Introduction}

Quantum circuit optimisation transforms a quantum circuit to improve the quality of its output. This quality can, for example, be measured by the fidelity of the output state to the ideal output state, or the accuracy of some derived quantity such as an observable expectation value. 
Often the number of two-qubit gates in the circuit is the dominant noise source \cite{QuantinuumH2Noise}, and so can be used as a proxy for the total noise affecting the circuit. In such cases, to increase the quality of a circuit's output it is desirable to reduce the number of two-qubit gates. 

\emph{Optimisation passes} for reducing two-qubit gate count are the most relevant for existing devices, and as such are abundant. Examples include passes for Clifford subcircuit optimisation \cite{Fagan_2019}, unitary resynthesis \cite{doecode_58510}, ZX-calculus-based rewriting \cite{Kissinger_2020}, and Pauli-gadget-based optimisation \cite{paykin2023pcoastpaulibasedquantumcircuit}. Many quantum software development kits, notably PyTKET \cite{Sivarajah_2021} and qiskit \cite{javadiabhari2024quantumcomputingqiskit}, offer default heuristic based sequences of optimisation passes, constructed to perform well in the typical case.

There are however at least two obstacles to the widespread effective utilisation of optimisation passes. Firstly, optimisation passes regularly target particular circuit structures or employ advanced circuit rewriting techniques. As such, recognising when an optimisation pass can be beneficially applied often requires extraordinary expertise in circuit construction and optimisation. Secondly, selecting a sequence of optimisations is susceptible to the \emph{Phase Ordering Problem} \cite{phase_ordering}, wherein the order of application of optimisations can impact the quality of the resulting circuit. Again, this can be hard to account for without very careful study of the particular circuit of concern. For the above reasons, fixed default pass sequences cannot be optimal in all cases, and a sequence bespoke to the circuit is preferred. 

Here we resolve this by using reinforcement learning (RL) to train an agent to analyse a quantum circuit, and propose a bespoke sequence of optimisation passes. We use Proximal Policy Optimisation (PPO) \cite{Schulman2017} with graph neural network (GNN) layers \cite{4700287}, leveraging the native graph structure of quantum circuits. The model draws actions from a set of existing optimisation passes, improving on the utilisation of already available techniques. In particular, we demonstrate that when our model chooses from a selection of PyTKET \cite{Sivarajah_2021} passes it outperforms default sequences of optimisation passes provided by PyTKET. On our comprehensive test set, the (mean, median) fraction of initial two-qubit gates removed by the agent is $(57.7\%, \ 56.7 \%)$, compared to $(41.8 \%, \ 50.0 \%)$ for the next best PyTKET default pass sequence. Finally, we compare our model to search-based approaches, which also create circuit-bespoke pass sequences. We demonstrate that our model produces optimised circuits of comparable two-qubit gate count to these search-based approaches, but at a notably lower time cost at execution.\footnote{This work is the subject of US patent application reference 63/966,606 QUK226911-US.}

\section{Related Work}

\textcite{Quetschlich_2025} consider the related problem of selecting the device and the default optimisation-pass sequence best suited for a particular circuit. Here we focus on outperforming the default passes and fix the device. Further, the choice of neural network architecture used in \textcite{Quetschlich_2025} limits the agent to inputs from a fixed-size observation space consisting of a fixed number of features of the circuit. The agent introduced in this work takes the whole circuit as input, with no limit to the size of the circuit imposed by the model architecture, so it is not limited to making predictions based on only a few features. \textcite{dangwal2025} also consider optimisation-pass selection but, rather than using neural networks, rely on classical simulation to rank optimisation passes. Clifford circuits similar to the original circuit, rather than the original circuit itself, are optimised and used to predict the performance of optimisation passes on the original circuit. Not using the original circuit limits the accuracy of the predictions, which is why we do use the original circuit here. 

This work sets out to improve the utilisation of existing \emph{global} optimisation passes; those which rewrite the entirety of, or large parts of, a circuit at once. We consider global passes as they are the most abundant and well developed, and have been composed into the most carefully considered baseline default pass sequences. However, other works have used machine learning (ML) to support \emph{local} rewriting. 

\textcite{fosel2021} employ convolutional neural networks acting on 3D grid representations of quantum circuits. The trained RL agent selects simple local rewrites, such as gate commutation and cancellation, on circuits with nearest neighbour connectivity. \textcite{li2023} utilise learned local representations of gates in a circuit to select a location in a circuit and a local transform to apply at that location, introducing a splitting of the action space to reduce its dimension. 

Neural networks have been used to select more complex local transformations based on the ZX calculus. \textcite{N_gele_2024} use GNNs to select local rewrites in order to reduce the node count of ZX graphs, while \textcite{Riu_2025} take a similar approach but to reduce two-qubit gate count in the extracted circuit. \textcite{charton2023} instead use transformers to learn existing ZX-based optimisation passes, while \textcite{mattick2025} use a GNN to guide a tree search over ZX rewrites. % Relatedly, \textcite{xu2022quartzsuperoptimizationquantumcircuits, Xu_2023} perform a greedy search over local rewrites.

Besides two-qubit gate count optimisation, \textcite{ruiz2024} consider T-gate count reduction; translating circuits into a 3D ``signature tensor'' and utilising an extension of the AlphaTensor~\cite{fawzi2022} transformer-based model to optimise portions of a circuit. Other works have considered the problems of \emph{routing} and \emph{circuit layout}, which take into account the constraints inherent to the connectivity of underlying hardware architecture.
\textcite{Zhou_2020} use Monte-Carlo tree search to achieve this, while \textcite{pozzi2020} is an early example of using RL for this task. As well as optimising existing circuits, ML has also been employed for the purpose of \emph{circuit synthesis} where a target state or unitary is given and a new circuit implementing it must be constructed, with \textcite{weiden2023} being one such example of this.
ML has also been employed to search for variational quantum algorithm ansatzes \cite{sadhu2024quantum}.

\section{RL for Pass Sequence Selection}

\begin{figure*}
    \centering
    \begin{subfigure}[c]{0.6 \textwidth}
        \includegraphics[scale=1.5]{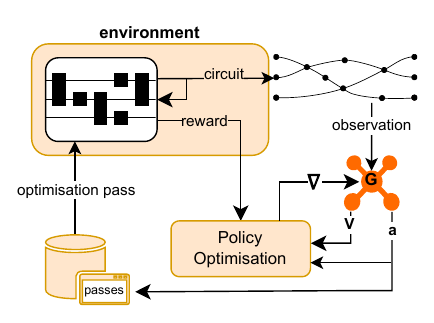}
        \caption{Training.}
        \label{fig:overview training}
    \end{subfigure}
    \begin{subfigure}[c]{0.3 \textwidth}
        \includegraphics[scale=1.5]{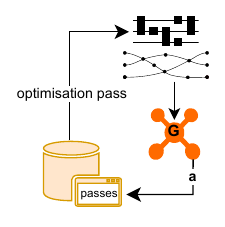}
        \caption{Deployment.}
        \label{fig:overview deployment}
    \end{subfigure}
    \caption{
        \textbf{Training and deploying an RL agent for optimisation-pass selection as described in this work.}
        \\
        \subref{fig:overview training}: A quantum circuit is loaded into the environment and transformed into a graph representation, as discussed in \cref{sec:graph representation}. This graph (the RL ``observation'' \cite{stable-baselines3}) is processed by a pair of GNNs, \textbf{G}, as discussed in \cref{sec:training_methodology}. The actor network outputs a vector of logits, \textbf{a}, for selecting the next optimisation pass; the critic head estimates the current circuit state value, \textbf{V}. An optimisation pass, as discussed in \cref{sec:action space}, is selected via softmax sampling of \textbf{a} and applied to the circuit. The reward, along with \textbf{V} and \textbf{a}, is used to compute gradients, $\nabla$,  and update \textbf{G} via PPO. The environment is reset with a fresh circuit from the training dataset (discussed in \cref{sec:training data}) whenever either: the \textsf{DoNothing} pass is selected; the two-qubit gate count is reduced to 0; or a preset number of passes are applied with no observed improvement.
        \\
        \subref{fig:overview deployment}: An input circuit is converted to a graph and repeatedly ingested by the actor network of \textbf{G}. Optimisation passes are selected via argmax sampling of \textbf{a}. The cycle terminates when the \textsf{DoNothing} pass is selected, and the optimised circuit is returned to the user.
    }
    \label{fig:overview}
\end{figure*}

In this work, we train an RL agent to, given a quantum circuit, iteratively select and apply a sequence of optimisation passes. The circuit is first encoded in a graph. PPO is used to train an actor-critic pair of GNNs to produce the best next pass in the sequence. \Cref{fig:overview} gives an overview of our method, which we describe in detail below.

\subsection{Graph Representation of Circuits}
\label{sec:graph representation}

Our graph encoding of the circuit uses nodes to represent gates, qubit inputs, and qubit outputs. Qubits are represented by edges, which give the order in which gates are applied to a given qubit. This allows for a natural representation of circuits with arbitrary connectivity, unlike 3D tensor representations of circuits. An example of the graph representation is given in \cref{fig:graph representation}.

\begin{figure}[t]
    \centering
    \begin{subfigure}{\linewidth}
        \centering
        \begin{tikzpicture}
    
            \draw[thick] (0.5,0.5) -- (1,0.5);
            \draw[thick] (0.5,1.5) -- (3,1.5);
            \draw[thick] (0.5,2.5) -- (8,2.5);
    
            \draw[thick] (1,0) rectangle (2.5,1) node[pos=.5]{$\mathsf{Rz \left( \gamma \right)}$};
    
            \filldraw[thick] (2,1.5) circle (3pt);
            \filldraw[thick] (2,2.5) circle (3pt);
            \draw[thick] (2,1.5) -- (2,2.5) node [midway, left] {$\epsilon$};
    
            \draw[thick] (3,1) rectangle (4,2) node[pos=.5]{$\mathsf{S}$};
    
            \draw[thick] (2.5,0.5) -- (8,0.5);
            \draw[thick] (4,1.5) -- (5,1.5);
    
            \filldraw[thick] (4.5,1.5) circle (3pt);
            \filldraw[thick] (4.5,0.5) circle (3pt);
            \draw[thick] (4.5,0.5) -- (4.5,1.5) node [midway, left] {$\delta$};
    
            \draw[thick] (5,1) rectangle (7.5,2) node[pos=.5]{$\mathsf{PhasedX \left(\alpha, \beta \right)}$};
    
            \draw[thick] (7.5,1.5) -- (8,1.5);
            
        \end{tikzpicture}
        \caption{An example circuit, comprising of \textsf{Rz}, \textsf{PhasedX}, and \textsf{ZZPhase} gates. \textsf{ZZPhase} gates are represented by black dots connected by vertical lines, with Greek letters (in this case $\delta$ and $\epsilon$) giving the associated gate angle.}
        \label{fig:cliff sub circuit circuit}
    \end{subfigure}
    \par\bigskip
    \begin{subfigure}{\linewidth}
        \centering
        \begin{tikzpicture}

            \definecolor{rx}{HTML}{DC3220}
            \definecolor{rz}{HTML}{005AB5}
            \definecolor{zz}{HTML}{3bb273}
            \definecolor{cliff}{HTML}{ffbd00}

            \draw[thick] (0,2.5) to [bend left] (1,2);
            \draw[thick] (0,1.5) to [bend right] (1,2);
            \draw[thick] (0,0.5) to [bend left] (1,0.5);

            \draw[thick] (1,2) to [bend right] (2,1.5);
            \draw[thick] (1,2) to [bend left] (5,2.5);
            \node at (3,3.15) {$\left[ \left[ 0 , 0, 1 , 0 \right] , \left[ 1 , 0 , 0 , 0 \right]\right]$};

            \draw[thick] (1,0.5) to [bend right] (3,1);
            \draw[thick] (2,1.5) to [bend left] (3,1);

            \draw[thick] (3,1) to [bend left] (4,1.5);

            \draw[thick] (4,1.5) to [bend right] (5,1.5);
            \draw[thick] (3,1) to [bend right] (5,0.5);

            \filldraw [zz] (1,2) circle (5pt) node [above=3pt] {$\epsilon$};
            \filldraw [rz] (1,0.5) circle (5pt)  node [above=3pt] {$\gamma$};
            
            \filldraw [color=cliff, fill=rz, very thick] (2,1.5) circle (5pt)  node [above=3pt] {$\frac{\pi}{2}$};

            \filldraw [zz] (3,1) circle (5pt)  node [above=3pt] {$\delta$};
            \filldraw [rx] (4,1.5) circle (5pt) node [above=3pt] {$\alpha,\beta$};

            % Inputs
            \filldraw [color=gray, fill=white, ultra thick] (0,0.5) circle (3pt);
            \filldraw [color=gray, fill=white, ultra thick] (0,1.5) circle (3pt);
            \filldraw [color=gray, fill=white, ultra thick] (0,2.5) circle (3pt);

            % Outputs
            \filldraw [color=gray, fill=white, ultra thick] (5,0.5) circle (3pt);
            \filldraw [color=gray, fill=white, ultra thick] (5,1.5) circle (3pt);
            \filldraw [color=gray, fill=white, ultra thick] (5,2.5) circle (3pt);
            
        \end{tikzpicture}
        \caption{The circuit graph of \subref{fig:cliff sub circuit circuit}. Colours correspond to gate type (green = \textsf{ZZPhase}, blue = \textsf{Rz}, red = \textsf{PhasedX}), with a yellow ring depicting the additional flag in the case of Clifford gates. Inputs and outputs are grey. An example edge encoding is given for the neighbouring edge, which corresponds to $[[0,0,1,0],[1,0,0,0]]$ because the source node of the edge corresponds to the `first' qubit acted on by the \textsf{ZZPhase} gate, and the destination node corresponds to an output.}
        \label{fig:cliff sub circuit dag}
    \end{subfigure}
    \caption{\textbf{Example circuit encoded as a graph.}}
    \label{fig:graph representation}
\end{figure}

Here we will use circuits in the universal \emph{Quantinuum native gateset}: \textsf{ZZPhase}, \textsf{Rz}, \textsf{PhasedX} \cite{QuantinuumH2Noise}
\begin{align}
    \textsf{ZZPhase} \left( \alpha \right) &= e^{-\frac{1}{2} i \pi \alpha Z \otimes Z}, \\ 
    \textsf{Rz} \left( \alpha \right) &= e^{-\frac{1}{2} i \pi \alpha Z}, \\ 
    \textsf{PhasedX} \left( \alpha , \beta \right) &=\textsf{Rz} \left( \beta \right) \textsf{Rx} \left( \alpha \right) \textsf{Rz} \left( - \beta \right),
\end{align}
where $\textsf{Rx} \left( \alpha \right) = e^{-\frac{1}{2} i \pi \alpha X}$. Each node is represented by an 8-element feature vector consisting of: a boolean flag indicating if the node is an input; a boolean flag indicating if the node is an output; the normalised \textsf{Rz} gate parameter; the set of normalised \textsf{PhasedX} gate parameters; the normalised \textsf{ZZPhase} gate parameter; a boolean flag indicating if the gate is a Clifford gate; and an index of the most recent optimisation pass that was applied to the circuit, or -1 if no pass has been applied. Only one of the three gate parameters can be non-zero, ensuring the node corresponds uniquely to a gate. Similarly a node cannot be a gate and an input/output, nor can a node be an input and an output.

The Clifford-gate flag is included to mitigate a representation issue with neural networks caused by Clifford angles being arbitrarily close to non-Clifford angles. The index of the most recent optimisation pass is included to avoid repetition of idempotent optimisation passes.

Edges have a source and a destination node. In the context of each node, the qubit corresponding to the edge can be either: an input/output qubit; the target of a single-qubit gate; the `first' qubit acted on by a two-qubit gate; or the `second' qubit acted on by a two-qubit gate.\footnote{$\textsf{ZZPhase}$ is symmetric so does not have a fundamental notion of `first' and `second'. Instead we assign these labels to qubits arbitrarily.}\footnote{Qubits are labelled as `first' and `second' in order to match input wires of the node to output wires when they concern the same qubit.} As such, the source and destination can each be described by a length-4 one-hot encoding, giving a length-8 vector for each edge. We use the above order of features in the example edge encoding of \cref{fig:cliff sub circuit dag}.

\subsection{Action Space}
\label{sec:action space}
 
The agent's action space consists of the following optimisation passes:
\begin{description}
    \item[\textsf{KAKDecomposition}:] Find and optimally resynthesise two-qubit subcircuits \cite{tucci2005introductioncartanskakdecomposition}.
    \item[\textsf{CliffordResynthesis}:] Find and resynthesise Clifford subcircuits \cite{Aaronson_2004}.
    \item[\textsf{CliffordSimp}:] Apply rewrite rules simplifying Clifford-gate sequences according to \textcite{Fagan_2019}.
    \item[\textsf{ZXGraphlikeOptimisation}:] Simplify circuit in ZX-calculus and extract circuit from simplified diagram according to \textcite{Backens2021therebackagain}.
    \item[\textsf{GreedyPauliSimp}:] Convert a circuit into a graph of Pauli gadgets and resynthesise according to \textcite{schmitz2023graphoptimizationperspectivelowdepth, paykin2023pcoastpaulibasedquantumcircuit}.
    \item[\textsf{ThreeQubitSquash}:] Find and resynthesise three-qubit subcircuits \cite{Shende_2006}.
    \item[\textsf{DoNothing}:] Do not alter the circuit and terminate.
\end{description}
Besides \textsf{\textsf{DoNothing}}, these passes are built from passes in PyTKET (version 2.9.3) with the same name. As the output gateset used by each PyTKET pass is different, after the application of each pass the circuit is also rebased to the Quantinuum native gateset.

\subsection{Training Methodology}
\label{sec:training_methodology}

We use a GNN to compute a learned representation of the nodes of the graph. We use graph convolutional layers  to ensure the receptive field of the graph nodes encompasses a large enough neighbourhood to capture discriminative features. To incorporate edge features, we use the modified \textsf{GINConv} operator of \textcite{hu2020}. We use skip connections to avoid vanishing gradients and improve information flow \cite{He2016}. We use mean pooling to construct a graph-level representation vector, ensuring that the trained network can ingest circuits of different sizes. A simple feed-forward network is used to compute action logits and state values from the graph-level representation. In our PPO implementation, gradient updates are performed using the Adam optimiser \cite{kingma2017}.

Let a circuit have $n_0$ two-qubit gates before any optimisation has taken place. After the application of $t$ optimisation passes it has $n_t$ two-qubit gates. The reward for the $t^{\text{th}}$ action is
\begin{equation}\label{eq:reward_function}
    r_t = \frac{n_{t-1} - n_{t}}{n_0}.
\end{equation}
We will use \emph{reward} to refer to the reward from the application of a single optimisation pass.
We will use \emph{cumulative reward} to refer to the total reward received by a sequence of optimisation passes; i.e. $\sum_{t} r_t$. The cumulative reward can be interpreted as the fraction of two-qubit gates removed by a pass sequence.

The normalisation by the denominator in \cref{eq:reward_function} guarantees that the maximum attainable cumulative reward for any circuit is 1.0. This ensures equal consideration of large and small circuits.
During training, to encourage efficient use of optimisation passes, we additionally include a small penalty for taking an action other than \textsf{DoNothing}.

The hyperparameters of the PPO algorithm and the specific architecture of the GNN are selected via hyperparameter optimisation. Trial parameters are sampled based on the Tree-structured Parzen Estimator algorithm \cite{bergstra2011}.
An initial search over the most influential parameters is carried out over $50$ trials. For the PPO, these are: the learning rate, the action penalty,\footnote{In practice the action penalty could be set according to the trade-off between the cost of spending time and compute resources on circuit optimisation versus the benefit gained from removing two-qubit gates. In the absence of a well-defined metric for this trade-off, we model the action penalty as a hyperparameter and determine its value through hyperparameter optimisation.} the batch size, the number of training epochs, and the rollout length per environment. For the GNN, these are the number of graph convolutional layers and the hidden layer dimension. Following this, a second search over 50 trials is carried out, where the optimal parameters from the initial search are fixed, and some additional PPO hyperparameters of secondary influence are scanned. These are the discount factor, $\gamma$, and the Generalised Advantage Estimation bias-variance trade-off factor, $\text{GAE}(\lambda)$. All hyperparameters, including those not covered by the search, are given in \cref{tab:ppo_gnn_hyperparams}.

\begin{table}%[h!]
    \centering
    \begin{tabularx}{\linewidth}{Xs}
        \toprule
        \textbf{Parameter} & \textbf{Value} \\
        \toprule
        \multicolumn{2}{l}{\textbf{PPO}} \\
        \midrule
        Learning rate &  $3.36 \times 10^{-4}$\\
        Rollout length per environment & $128$ \\
        Batch size & $64$ \\
        Number of epochs & $3$ \\
        Discount factor ($\gamma$) & $0.952$ \\
        $\text{GAE}(\lambda)$ & $0.938$ \\
        Clip range & $0.2$ \\
        Entropy coefficient & $0.01$ \\
        Value function coefficient & $0.5$ \\ Gradient clipping maximum value & $0.5$ \\
        Action penalty & $0.013$ \\
        \toprule
        \multicolumn{2}{l}{\textbf{GNN Architecture}} \\
        \midrule
        Number of convolutional layers & 4 \\
        Hidden layer dimension & 128 \\
        \bottomrule
    \end{tabularx}
    \caption{\textbf{Optimised hyperparameters for the PPO and GNN architecture.}}
\label{tab:ppo_gnn_hyperparams}
\end{table}

\subsection{Datasets}
\label{sec:training data}

Training and testing is performed with synthetic circuits, randomly generated from the following circuit classes:
\begin{description}
    \item[Random-SU4:] Random $\text{SU}(4)$ gates acting on random qubit pairs. Inspired by quantum volume circuits \cite{Cross_2019}.
    \item[Random-SU8:] Random $\text{SU}(8)$ gates acting on random triples of qubits.
    \item[IQP:] Random sequences of commuting gates, introduced in \textcite{Shepherd2009} and central to quantum computational supremacy arguments \cite{Bremner_2016}. IQP circuits have since been used in quantum machine learning with two examples being \textcite{Havl_ek_2019, Coyle_2020}.
    \item[QAOA:] Random instances of a popular ansatz used for optimisation problems \cite{farhi2014}.
    \item[Pauli:] Random sequences of Pauli gadgets \cite{Cowtan_2020}.
    \item[Clifford-SU4:] Circuits composed of a mixture of random Clifford subcircuits and random $\text{SU}(4)$ gates.
    \item[Ordered-Clifford-SU4:] Constructed by sandwiching random Clifford identities between sets of random two-qubit gates.
    \item[Clifford-SU4-SU8:] Circuits composed of a mixture of random Clifford subcircuits, random $\text{SU}(4)$ gates, and random $\text{SU}(8)$ gates.
\end{description}
These circuits are a mixture of application motivated circuits\footnote{Random-SU4, IQP, and Pauli are taken from the standard application motivated benchmarking set proposed by \textcite{Mills2021application}.} and circuits constructed to expose the model to complex structures. These circuits are expected to require relatively short sequences of the actions listed in \cref{sec:action space},\footnote{Circuits requiring long sequences of passes are rare. This is in part because global optimisation passes often conflict with each other when run in sequence. This is an important aspect of the Phase Ordering Problem, and is discussed further in \cref{sec:conclusion}.} but recognising the correct sequence by eye is challenging and time consuming. We minimally pre-process the circuits so that they are in the Quantinuum default gateset. To thoroughly test the optimisation methods, we do not assume that any prior optimisation has occurred. We do not impose any connectivity restrictions on the circuits we generate, which is naturally supported by our use of GNNs.

The training set comprises 402,558 circuits, of which 10\% are withheld for validation. Circuits are of a range of sizes, with qubit counts spanning the range $[3, 7]$, and two-qubit gate counts spanning the range $[0, 215]$.\footnote{For circuits generated with no two-qubit gates, the model performs no optimisation and simply returns the original circuit, giving a total reward of 0.}

\subsection{Training and Deployment}
\label{sec:training and deployment}

The model is trained using PPO. PPO is an established and well studied approach whose clipped objective is designed to limit excessive policy changes during training. PPO supports our discrete action space, and provides a practical baseline for evaluating learned pass selection. As we will see in \cref{sec:results}, in practice training with PPO is stable and gives strong results. We find the training resources required by PPO are manageable, while the trained model generalises well to large circuits. In this work we will show that PPO can be used effectively to meet our use case, not that it is necessarily the most efficient tool for doing so. We leave this broader analysis to future work.

The PPO algorithm we use is as implemented in Stable Baselines3 \cite{stable-baselines3} (version 2.7.1), employing vectorised execution over {\color{red}6} parallel environment instances. In this implementation, the PPO rollout length is the number of environment transitions collected from each parallel instance before a policy update. Thus, the value 128 in \cref{tab:ppo_gnn_hyperparams} produces a rollout buffer of $128 \times 6 = 768$ transitions per update. These transitions may span many optimisation episodes and circuits; 128 is not the number of optimisation passes applied to one circuit, nor a fixed deployment-time pass-sequence length. Experiments are conducted on a Lightning-AI-hosted \cite{lightningai} machine equipped with a single NVIDIA L4 Tensor Core GPU, 16 vCPUs, and 64 GB of RAM. Evaluation is also performed on the same machine, unless otherwise stated.

Training is limited to 300,000 environment transitions (each transition applies one selected pass in one training environment), with early stopping triggered when the performance on validation data stagnates. The learning curves of validation cumulative reward against environment transitions for all the trained seeds are presented and discussed in \cref{app:validation learning curve}. For all seeds, learning saturated before $300,000$ transitions, with early stopping typically triggered at fewer than $150,000$ environment transitions. This is encouraging, as, while sampling the environment is costly, training typically stops early, and in a reasonable amount of time.

For an independent ``timing'' run, early stopping occurred after $135,168$ environment transitions, corresponding to a wall-clock training time of 2.98 hours. Of this total wall-clock training time: approximately 79.85\% was spent performing environment stepping, 8.93\% was spent performing gradient updates, and 11.22\% was spent performing other tasks. This demonstrates that the time to train the model is dominated by obtaining samples, which is to say, by running the optimisation passes. This breakdown of the wall-clock training time suggests that an alternative to PPO with lower sampling requirements -- such as off-policy methods like DQN \cite{mnih2015humanlevel} -- could accelerate training.

Once the model has been trained, the actor network can be easily deployed. An input circuit, represented in graph form, is ingested by the network, which selects the next pass which is then applied to the circuit. This process is repeated until one of the following occurs: the two-qubit gate count is $0$; the \textsf{DoNothing} action is selected; a fixed number (configurable) of passes has been applied with no further reduction in two-qubit gate count. At this point, the optimised circuit is returned to the user.

\section{Results}\label{sec:results}

Here we present results comparing our trained model to the following default PyTKET (version 2.9.3) optimisation-pass sequences:
\begin{description}
    \item[\textsf{FullPeepholeOptimise}:] A general-purpose sequence of 6 passes (not including rebasing) performing a combination of one-, two-, and three-qubit subcircuit squashing, and Clifford circuit simplification.
    \item[\textsf{QuantinuumDefaultTwo}:] Extends \textsf{FullPeepholeOptimise} with very lightweight one- and two- qubit squashing passes to optimise for the Quantinuum H-series devices.
    \item[\textsf{QuantinuumDefaultThree}:] Extends \textsf{GreedyPauliSimp} with very lightweight one- and two- qubit squashing passes to optimise for the Quantinuum H-series devices.
\end{description}
\textsf{FullPeepholeOptimise} is derived from a PyTKET pass of the same name. The latter two passes are default optimisation passes for the PyTKET Quantinuum backend (optimisation levels set as in the pass name \cite{pytketquantinuum}). Note that \textsf{QuantinuumDefaultTwo}, \textsf{QuantinuumDefaultThree}, and \textsf{FullPeepholeOptimise} are built from several individual passes as detailed in \textcite{Sivarajah_2021}. We additionally compare to \textsf{GreedyPauliSimp} and \textsf{KAKDecomposition}, as these are particularly performant single passes.

\paragraph{In-Distribution Testing}

\begin{figure*}[t]
    \centering
    \includegraphics[width=0.9\linewidth]{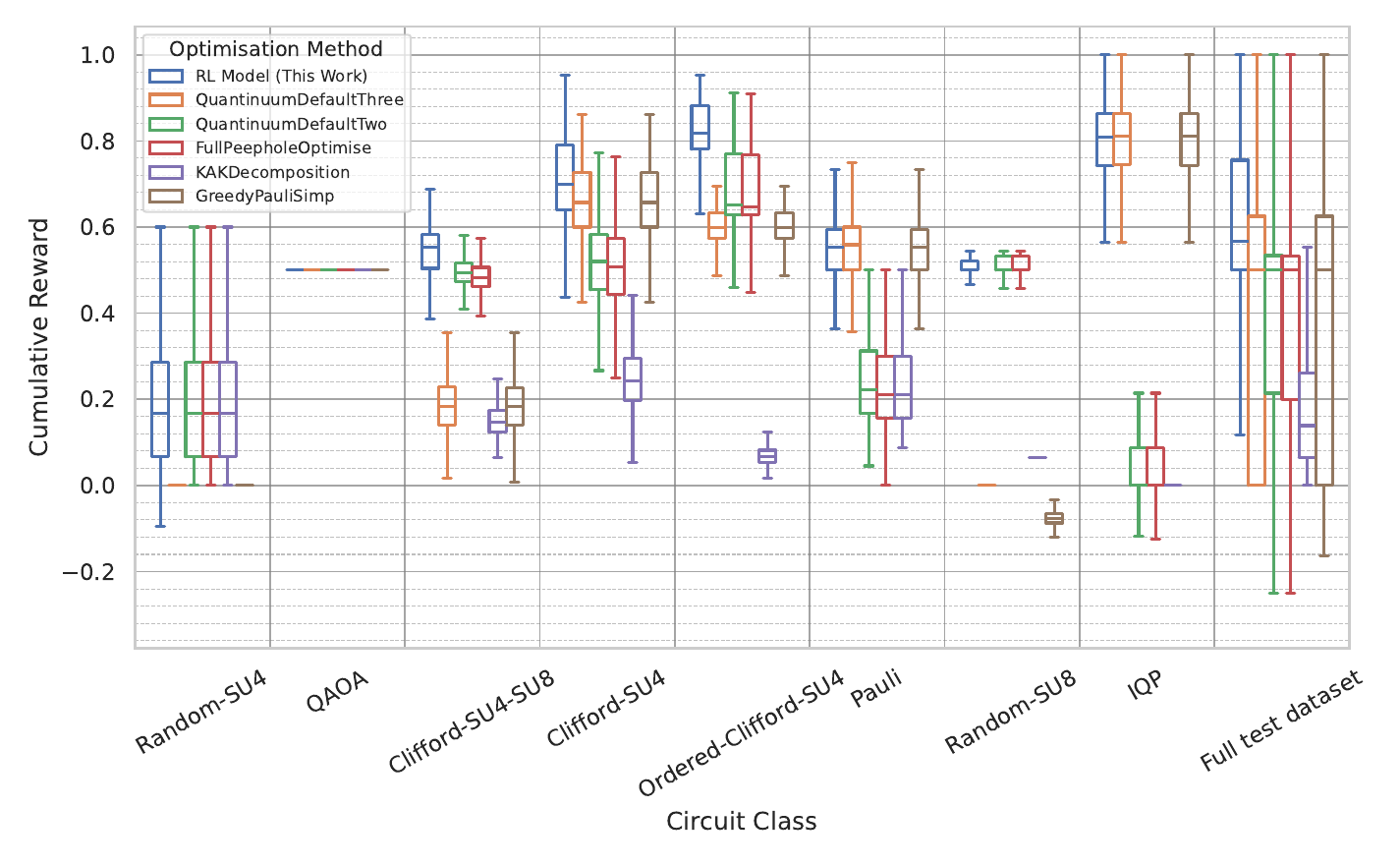}
    \caption{\textbf{Cumulative reward (fraction of two-qubit gates removed) for in-distribution circuits.}
    $40,412$ circuits are optimised with nine trained models and a selection of PyTKET optimisation passes. Box plots give the median and interquartile range. Whiskers extend to the lowest (highest) data point still within $1.5$ times the interquartile range of the lower (upper) quartile. `Full test dataset' accumulates all of the circuits.} 
    \label{fig:results_chart}
\end{figure*}

Results in \cref{fig:results_chart} report performance on a test dataset consisting of 40,412 circuits, approximately equally distributed across circuit classes. These circuits are unseen during training, but are drawn from the same distribution as those in the training set; hence referred to as `in-distribution' circuits. The model is trained 9 times, each time with a different initial seed (weights initialisation), to give an agent for each seed.\footnote{Due to stochastic training, performance varies across random seeds. We report statistics over multiple independently trained agents to assess robustness and avoid bias from selecting a single best-performing run.} Each circuit is optimised once with each agent. The baseline optimisation passes are deterministic, therefore each circuit is optimised only once per baseline optimisation pass. As such, in total there are $40{,}412 \times (9 + 5) = 565{,}768$ optimised circuits; 40,412 for each baseline default pass, and 363,708 for the `RL Model'. The box plots in \cref{fig:results_chart} give the distribution, over optimised circuits, of the cumulative reward. 

The median cumulative rewards over all circuits (`Full test dataset' in \cref{fig:results_chart}) are 0.567 and 0.500 for the model and the best PyTKET default pass respectively (\textsf{QuantinuumDefaultTwo}). Further, the distribution of cumulative reward for the model is skewed upwards as compared to the other passes. This is reflected in the notable improvement in mean cumulative reward: 0.577 for the model, compared to 0.418 for \textsf{QuantinuumDefaultTwo}.
This demonstrates that while the default passes are typically good, a bespoke approach is preferred in many cases. In particular, most default passes have a circuit class or two where they perform particularly poorly. Conversely, for the majority of circuit classes the model generates the sequence of passes with the highest, or close to the highest, median cumulative reward of all the passes compared.

This skew in the distribution is significant in practice, as it demonstrates that the worst performance a user can expect when using the model presented is still better than the worst performance they can expect from any existing pass sequence. Similarly, the best performance they can expect is better than for any particular existing pass sequence. This is important, as in practice \emph{any} circuit that a user runs is valuable to them, not just the ``average'' circuit.

\begin{figure*}[t]
    \centering
    \includegraphics[width=0.9\linewidth]{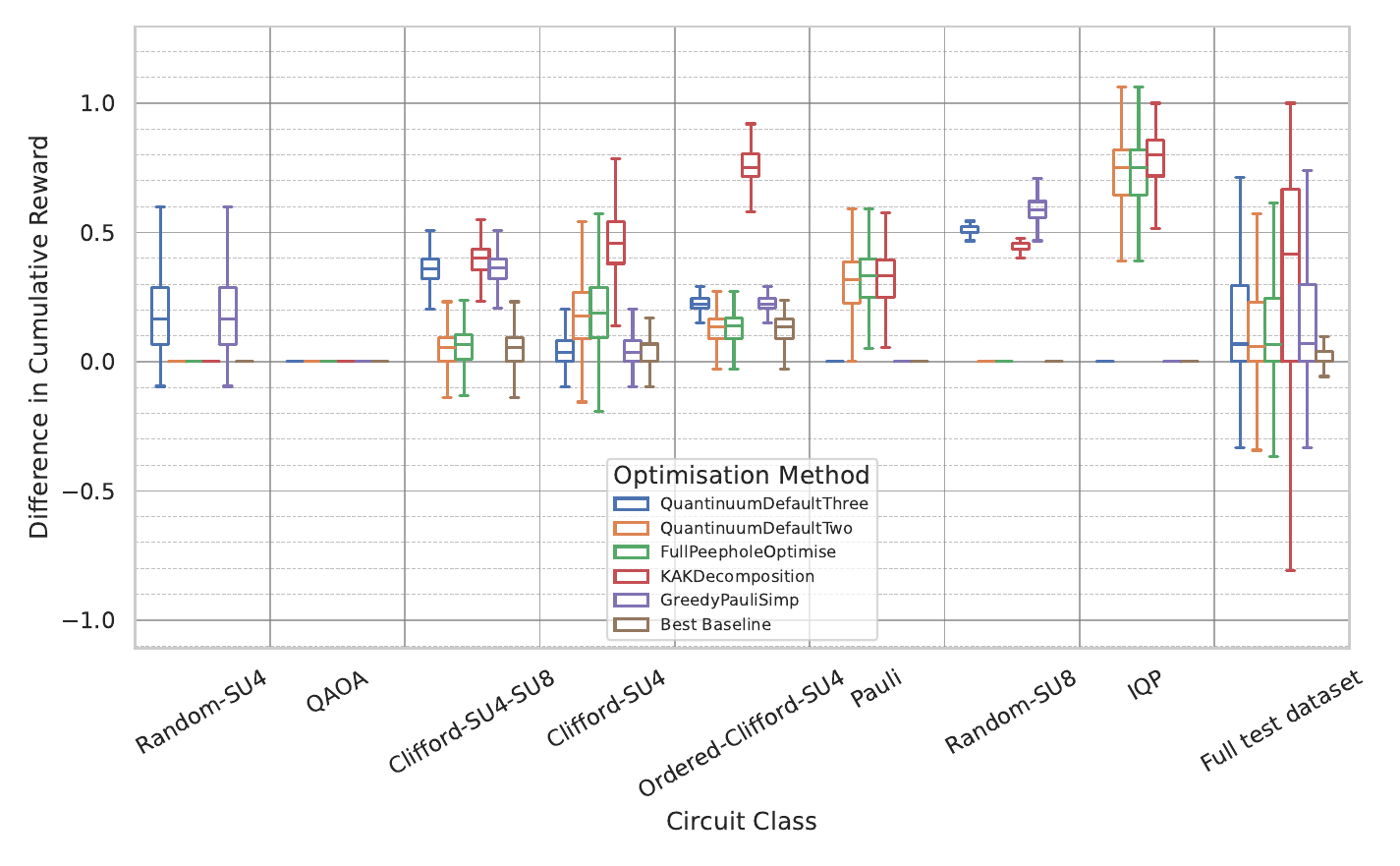}
    \caption{\textbf{Difference in cumulative reward (fraction of two-qubit gates removed) between the RL model and baseline optimisation passes for in-distribution circuits.} 
    Positive values indicate improved performance of the RL model. 40,412 circuits are optimised with nine trained models and a selection of PyTKET optimisation methods. Box plots give the median and interquartile range. Whiskers extend to the lowest (highest) data point still within $1.5$ times the interquartile range of the lower (upper) quartile. `Full test dataset' accumulates all of the circuits. `Best Baseline' corresponds to the highest cumulative reward generated by any baseline approach for each circuit.} 
    \label{fig:results_improvement_chart}
\end{figure*}

\Cref{fig:results_improvement_chart} gives the distributions, over optimised circuits, of the difference between the model's cumulative reward and the cumulative reward of the baseline methods. This shows that the median performance of the model is always as good as -- or better than -- the baseline methods. We also see that for the Clifford-SU4-SU8, Clifford-SU4, and Ordered-SU4-SU4 circuit classes, the median performance is always better than even the best baseline method. This again demonstrates that the model does more than simply reproduce the best baseline approach.

\begin{figure}
    \centering
    \includegraphics[width=\linewidth]{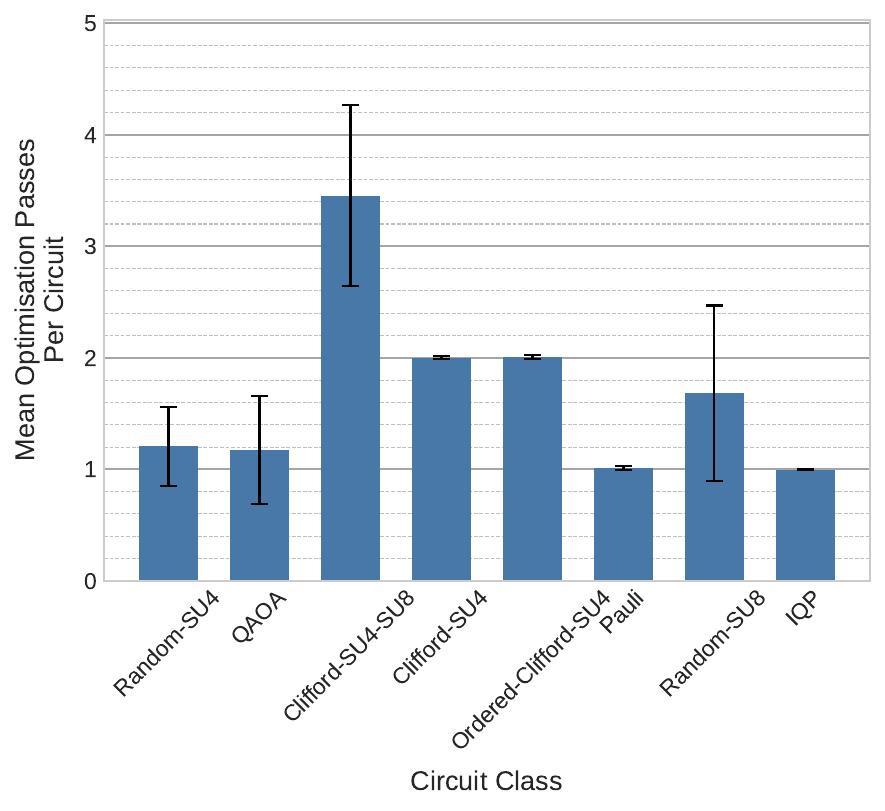}
    \caption{\textbf{Mean number of optimisation passes selected per circuit for each circuit class.} Bars show the mean of the per-run averages across the nine PPO training runs, and error bars indicate $\pm 1$ sample standard deviation across those averages. Counts include all optimisation passes in each recorded evaluation sequence, excluding the \texttt{DoNothing} action.} 
    \label{fig:average_num_passes}
\end{figure}

It is important to note that the agent is indeed generating a bespoke sequence of passes for each circuit, rather than, say, simply applying all passes repeatedly. This is important for the usability of the trained agent, as the time taken to execute excessively long sequences of passes would be very costly. For example, the model performs identically to the \textsf{GreedyPauliSimp} in the case of IQP and Pauli circuits. This is because just a single pass is best in this case, with the agent performing a one-pass sequence consisting only of \textsf{GreedyPauliSimp} in $\sim 99.6 \%$ and $\sim 99.1 \%$ of cases, respectively. In the case of Clifford-SU4 and Clifford-SU4-SU8, a longer sequence of passes can be required as these circuits consist of a more complicated mixture of gates. In particular, for Clifford-SU4-SU8 the average length of the pass sequence identified by the model is $\sim 3.36$, excluding the \textsf{DoNothing} action. In the case of these two circuit classes, the agent outperforms existing default sequences of passes, suggesting that general purpose passes are not well suited here. The average number of passes chosen by the model across all seeds is given in \cref{fig:average_num_passes}.

In the case of Ordered-Clifford-SU4 circuits, the order of application of passes is important. In particular, removing the Clifford identities with \textsf{CliffordSimp} or \textsf{GreedyPauliSimp} allows the remaining two-qubit subcircuits to be optimised with \textsf{KAKDecomposition}. The model is almost always (in $99.2\%$ of cases) able to identify this as the correct sequence of passes. Note that neither \textsf{CliffordSimp} or \textsf{GreedyPauliSimp} are the best single first pass to apply, and if only one pass was possible then \textsf{ThreeQubitSquash} would perform better. This demonstrates the model's ability to delay reward in order to select the best sequence of passes.

\paragraph{Generalisability}

\begin{figure*}[t]
    \centering
    \includegraphics[width=0.9\linewidth]{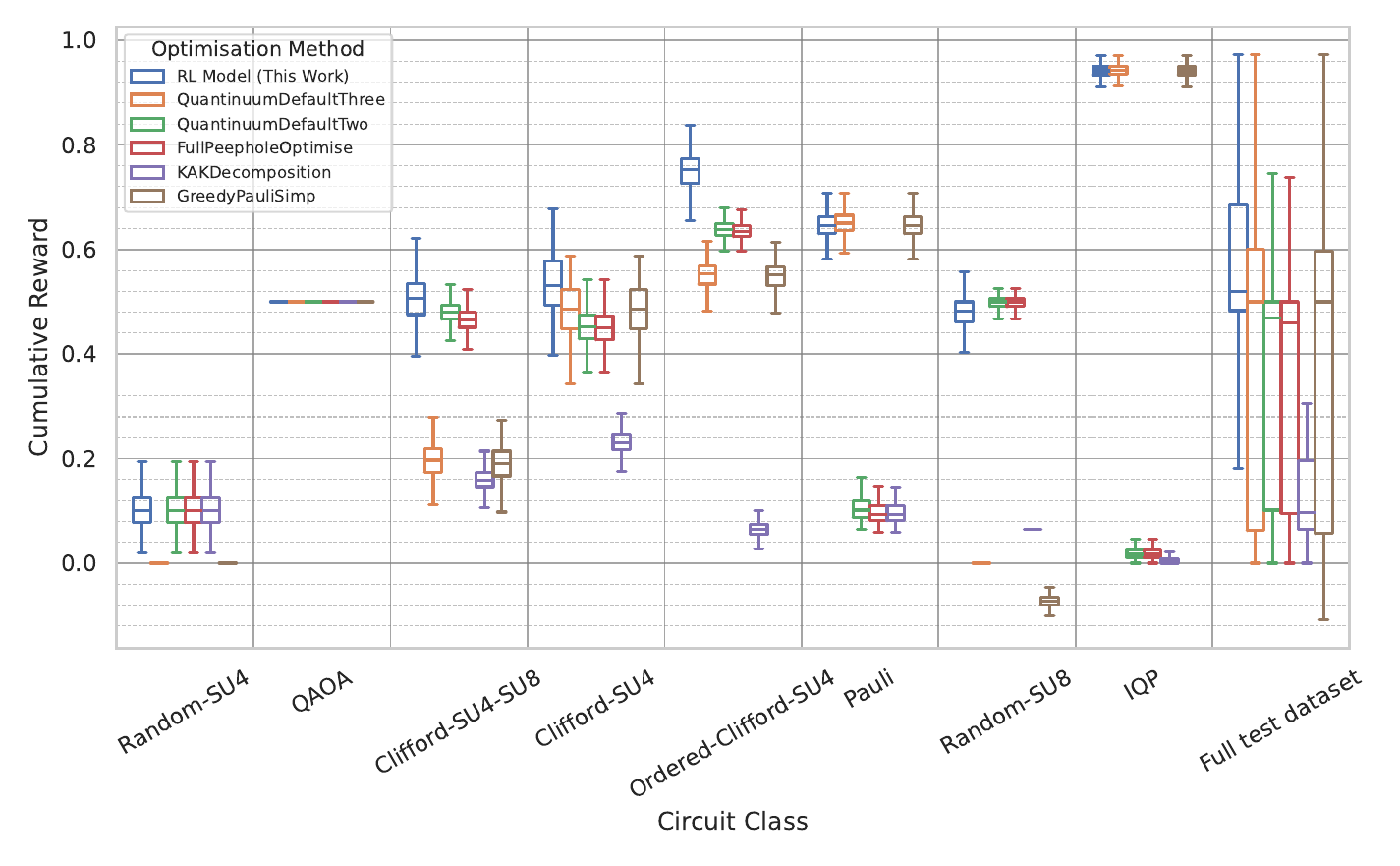}
    \caption{\textbf{Cumulative reward (fraction of two-qubit gates removed) for out-of-distribution circuits.} 8,115 circuits are optimised with one trained model and a selection of PyTKET optimisation passes. Box plots give the median and interquartile range. Whiskers extend to the lowest (highest) data point still within $1.5$ times the interquartile range of the lower (upper) quartile. `Full test dataset' accumulates all of the circuits.} 
    \label{fig:generalisation}
\end{figure*}

\Cref{fig:generalisation} compares performance on a test set of 8,115 circuits, roughly equally spread across circuit classes, which are larger than anything in the training set; hence referred to as `out-of-distribution' circuits. Qubit count is in the range $[8, 12]$, and two-qubit gate counts span $[216, 450]$. As the individual optimisation passes take longer to run on larger circuits, we test with fewer circuits here than for the in-distribution testing, and use just one of the nine trained models. As we do not observe significant variation between seeds, one seed is chosen at random. We see that performance in \cref{fig:generalisation} is comparable to that seen in \cref{fig:results_chart}: the (mean, median) fraction of two-qubit gates removed by the agent is $(55.7\%, \ 52.0 \%)$, compared to $(41.5 \%, \ 50.0 \%)$ for the next best default pass sequence (\textsf{QuantinuumDefaultThree}). This is encouraging, and demonstrates that the model generalises to circuits larger than those seen in the training data. Importantly, this demonstrates that we can deploy the model on large circuits while conducting very manageable training on smaller circuits.

\begin{figure}
    \centering
    \includegraphics[width=0.9\linewidth]{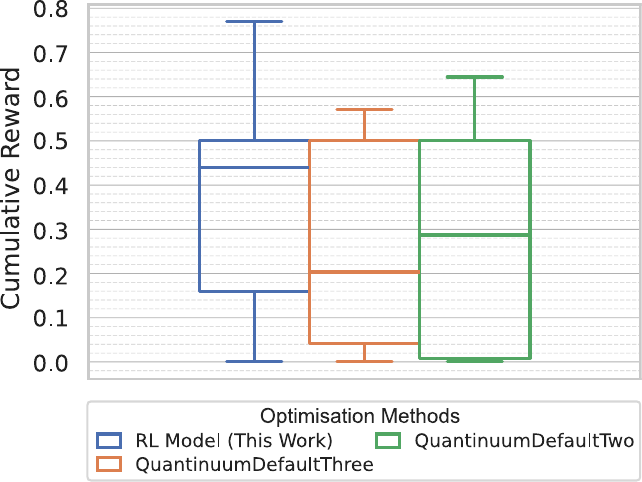}
    \caption{\textbf{Cumulative reward (fraction of two-qubit gates removed) for very large out-of-distribution circuits.} 1,000 circuits acting on 100 qubits, equally distributed between all circuit types in \cref{sec:training data}, are optimised with one trained model, \textsf{QuantinuumDefaultTwo}, and \textsf{QuantinuumDefaultThree}. Box plots give the median and interquartile range across all circuit types. Whiskers extend to the lowest (highest) data point still within $1.5$ times the interquartile range of the lower (upper) quartile.} 
    \label{fig:very large generalisation}
\end{figure}

In \cref{fig:very large generalisation} we present results from executing our model and the Quantinuum default passes on 1000 very large circuits, with 100 qubits and two-qubit gate counts in the range $[ 878 , 1150 ]$. This is slightly beyond the largest all-to-all connected trapped-ion architectures demonstrated to date~\cite{ransford2025helios98qubittrappedionquantum}. Again, as executing passes in our action space takes longer with such large circuits, we use fewer than in \cref{fig:generalisation}, but observe similar relative performance between the optimisation passes. The absolute performance is reduced, likely because the circuits are wide and shallow as compared to the circuits used in \cref{fig:generalisation,fig:results_chart}.

While \cref{fig:generalisation,fig:very large generalisation} demonstrate what is possible with our approach, we note that we are limited in this regard by the passes in our action space. Since applying the optimisations constitutes the greatest bottleneck, with time to evaluate the model being minimal, improvements to the optimisation passes would increase the reach of our method.

\paragraph{Scalability}

Finally we compare the model to a simple beam-search-based optimiser \cite{Lowerre1976}. Beam search is selected as it too generates optimisation-pass sequences bespoke to a given circuit:
\begin{description}
    \item[\textsf{$\text{Depth}[n]\text{Width}[m]$}:] To construct the search tree, we take the unoptimised circuit to be the root node (level 0). Child nodes are constructed from parents by applying all optimisation passes in the action space, giving one child node per action. At each level of the tree, only a fixed number of nodes (the width, $m$) are retained, with the worst nodes pruned. This is repeated up to a fixed number of levels (the depth, $n$). The best leaf node is returned as the optimised circuit.
    \item[\textsf{GreedySearch}:] Apply each of the optimisation passes in the action space to the unoptimised circuit, and select the optimised circuit with the lowest two-qubit gate count. This can be thought of as a depth-1 beam search.
\end{description}
\textsf{$\text{Depth}[n]\text{Width}[m]$} generates a length-$n$ optimisation-pass sequence, while \textsf{GreedySearch} generates length-one sequences.

Here we are concerned with exploring the scalability of these approaches to larger circuits and action spaces, and to longer optimisation-pass sequences. As such, we consider circuits in the Clifford-SU4, Ordered-Clifford-SU4, and Clifford-SU4-SU8 circuit classes. These have the most complex structures, and typically require longer optimisation-pass sequences. 

In total, we optimise 930 circuits approximately distributed between these three classes; once with each of the search-based optimisation passes, and once with the model. Again, we use only one trained model; i.e. one seed value. Note that we use fewer circuits than in the plots above, to account for the greater runtime of the search-based methods. Qubit counts are in the range $[ 4 , 8 ]$, and two-qubit gate counts are in the range $[ 100 , 300 ]$.
These experiments were conducted on an Apple Macbook Pro with an M3 CPU and 36GB RAM.

\begin{figure}
    \centering
    \begin{subfigure}[t]{\linewidth}
        \centering
        \includegraphics[width=\linewidth]{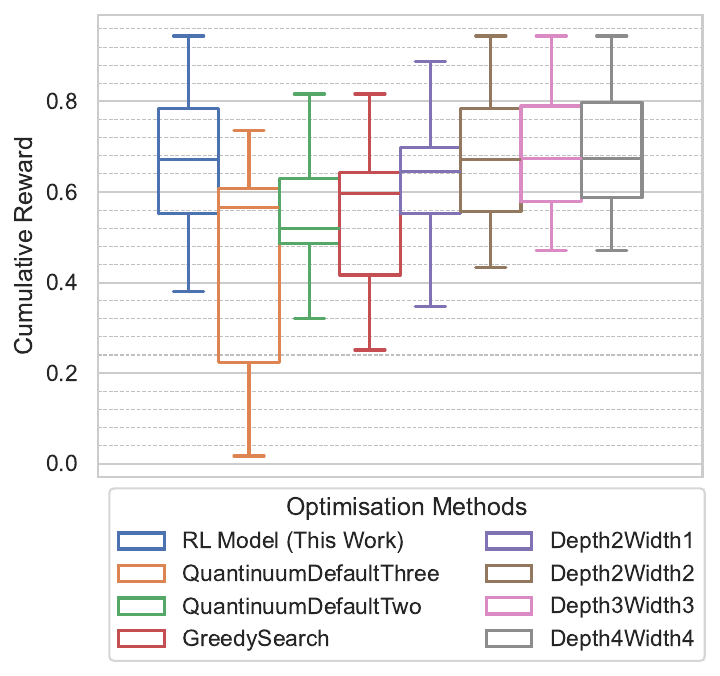}
        \caption{Comparison of cumulative reward (fraction of two-qubit gates removed). Box plots give the distribution across all circuits. Boxes give median and interquartile range, with whiskers extending to the largest/smallest values within 1.5 times the interquartile range.}
        \label{fig:beam_search_2q}
    \end{subfigure}
    \begin{subfigure}[t]{\linewidth}
        \centering
        \includegraphics[width=\linewidth]{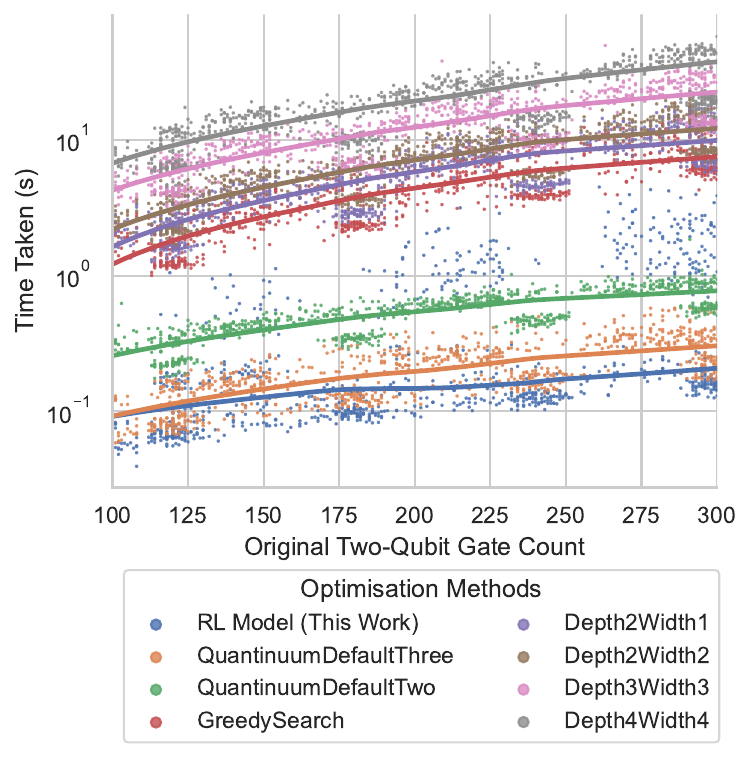}
        \caption{Comparison of time to optimise. Lines of best fit are a locally weighted scatter-plot smoothing.}
        \label{fig:beam_search_time}
    \end{subfigure}
    \caption{\textbf{Comparison to search-based optimisation techniques.} A total of 930 circuits from the Clifford-SU4, Ordered-Clifford-SU4, and Clifford-SU4-SU8 circuit classes are optimised for each of the optimisation methods.}
    \label{fig:beam_search}
\end{figure}

\Cref{fig:beam_search_2q} demonstrates that the model outperforms \textsf{GreedySearch} and \textsf{Depth2Width1}, and performs comparably to the remaining search-based approaches. However, \cref{fig:beam_search_time} shows that the model is notably faster than all of the search-based optimisation methods. This is true even for the very simple \textsf{GreedySearch}, which is comfortably outperformed by our model. We additionally see that the time taken by the model is comparable to \textsf{QuantinuumDefaultTwo} and \textsf{QuantinuumDefaultThree}, although the time taken by the model has a greater spread. This is because the model can select an arbitrary number of passes, while \textsf{QuantinuumDefaultTwo} and \textsf{QuantinuumDefaultThree} apply a fixed number. We see that the beam-search-based approaches improve in performance as the width and depth increases, but that performance plateaus, with the time taken continuing to increase.

This highlights several disadvantages of search-based approaches as compared to our RL-based approach. For a fixed circuit size, the time taken by beam-search-based approaches scales like $\text{action space size} \times \text{search width} \times \text{search depth}$. The time to evaluate the model is minimal compared to the time to execute the action selected. As such, the time taken by our model scales only with the depth. Thus, search-based approaches inevitably scale poorly as the action space grows. Additionally, when employing search-based approaches it is typically not clear at what level of the search tree to stop. In combination, these pitfalls validate the RL-based approach we have taken here.

It is important to note that for both the model- and search-based optimisers, parallelism can be utilised to decrease the runtime of deployment. We do not consider such improvements in either case in this comparison. Further, as we have discussed, the training time for the model can be notable: 2.98 hours in this case. However, the training need only occur once for the model to then be used as many times as is required. As such, we regard the analysis in this section to be representative of the performance of the model when deployed in practice.

\section{Conclusion}
\label{sec:conclusion}

We have trained an RL agent to generate, given a quantum circuit, a sequence of optimisation passes to best optimise that circuit. The trained agent markedly outperforms default PyTKET optimisation-pass sequences on average over our test set. For each class of circuits in the test set, the agent produces the best -- or close to the best -- sequence of optimisation passes. We have demonstrated the generalisability of the trained model to circuits larger than anything seen in the training data. We have shown that our RL-based approach is more scalable than other search-based approaches to building bespoke optimisation-pass sequences.

This agent reduces the expertise in quantum circuit compiler theory that is required to orchestrate PyTKET optimisation passes. To make the comparison to default PyTKET optimisation sequences fair, we have limited the action space to PyTKET passes. However, the action space can be straightforwardly expanded to include optimisation passes from other libraries, without any notable changes to the methodology described. We anticipate that an agent trained as described in this paper, but with this expanded action space, will perform better still.

Indeed, our approach clearly separates the problem of designing circuit optimisation passes from the problem of how to apply them in practice. We expect this separation of concerns to facilitate accelerated development in both fields. In particular, future work may look into new models and training protocols to improve optimisation-pass sequencing. In parallel, circuit optimisation research can proceed knowing that the passes developed will be optimally deployed.

In this article we have assumed arbitrary circuit connectivity, and have limited ourselves to the native Quantinuum gateset. Other circuit restrictions may be relevant to other devices. The gateset is hardcoded into: the graph encoding; the actions (which always preserve the gateset); and the reward, which counts \textsf{ZZPhase} gates. A different gateset would require minimal change to these hard-coded elements, and the overall methodology proposed in this paper would apply. Note that gatesets with, say, 3-qubit gates or larger -- such as Toffoli gates -- would not be problematic. In this article, we encode circuits in graphs of maximum degree 4; however, GNNs are flexible and can accommodate corresponding graphs of degree 6 or greater. Compilation barriers may also be straightforwardly encoded with their own node type, and would not cause a problem so long as the passes in the action space can be used safely on circuits with barriers. A different device connectivity can be accounted for by constructing actions that respect the device connectivity -- say by applying a routing pass at the end of each action. Limiting the actions by insisting that they preserve the gateset or circuit connectivity can miss opportunities for optimisation. Future work should explore designing a reward function which can allow this.

We have demonstrated that bespoke sequences of optimisation passes selected by the model outperform default sequences defined for general use. However using passes which perform global rewrites is still fundamentally limited by the Phase Ordering Problem; passes rewriting the whole circuit may improve the two-qubit gate count in one region of the circuit, but make it worse in another region. A natural application of RL extending the ideas in this paper would be to develop an agent that can select regions of the circuit to rewrite, in addition to selecting the pass to perform on just that region. Note that the region selected may be the whole circuit, combining the benefits of both global and local optimisation. This combines the benefits of our work with that of \textcite{li2023}, where the authors use an RL agent to select regions on which to apply rewrites (although the rewrites are much simpler than the global rewrites we explore here).

To further improve the performance of the agent it would be beneficial to expand our training and testing sets to include more circuit types and sizes. In particular, using a training set consisting wholly or partly of real user circuits would be beneficial, although we are not aware of any such sufficiently large training set available at present. On the topic of testing, while the standard application-motivated benchmark set presented in \textcite{Mills2021application} forms a subset of our testing set, in the future the model should be evaluated against other external standardised circuit collections, for example \textcite{quetschlich2023mqtbench}. The immediate obstacle to doing so is that gates and directives (especially barriers) outside of the gateset on which the current model is trained often feature in such circuit collections, as is the case for \textcite{quetschlich2023mqtbench} in particular.

Here we have focused on reducing two-qubit gate count; the dominant noise source for most existing QPUs, and the metric targeted by the majority of optimisation passes available. However, the dominant sources of logical errors when running fault-tolerant circuits may be otherwise; being, for example, $\textsf{T}$ gates when magic state distillation is required \cite{Bravyi_2005}. We expect similar techniques to those developed here to be beneficially applied to $\textsf{T}$-gate count reduction. We encourage investigations of that kind when a notable number of such passes are available. A second consideration for fault-tolerant circuits are the circuit types and sizes that will be of concern. We have demonstrated that our model works well, and can be trained with PPO with manageable resources, for circuits of sizes which can reasonably be run on existing and near-term devices. However, fault-tolerant circuits may have on the order of millions of gates, at which point the sample efficiency of the training may become a concern. To that end, more sample efficient alternatives to PPO, such as off-policy methods like DQN, should be investigated.

\paragraph{Data Availability Statement}
The data that support the findings of this study are openly available at the following URL/DOI: \url{https://doi.org/10.5281/zenodo.18786469} \cite{mills2026datasets}.

\paragraph{Code Availability}
The code used to generate the plots and reported results is available at \url{https://github.com/Quantinuum/rl-for-adaptive-circuit-optimisation} and uses the data deposited at the aforementioned Zenodo repository. The code used to define and train the models is not publicly available.

\paragraph{Acknowledgements}
The authors would like to thank Yao Tang and Matteo Puviani for their careful reviewing of the manuscript. The authors would also like to thank Alan Lawrence, Frederic Rapp, and Stephen Clark for insightful discussions.

\newpage

\printbibliography

\onecolumn

\appendix

\section{Validation Learning Curves}
\label{app:validation learning curve}

\begin{figure*}[ht!]
    \centering
    \includegraphics[width=0.9\linewidth]{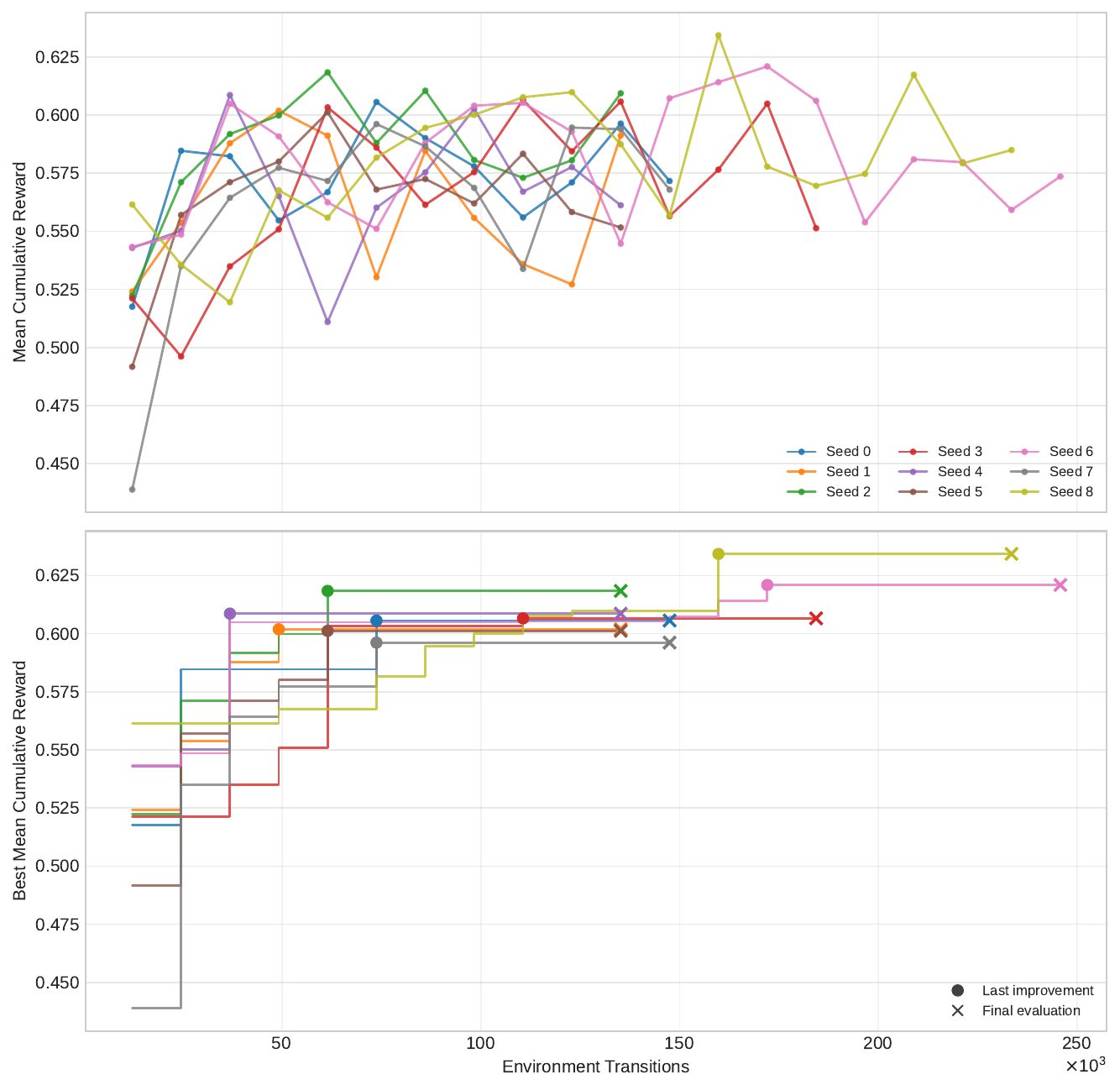}
    \caption{\textbf{Mean validation learning curves on in-distribution circuits for the 9 PPO training runs.} The top panel shows the mean validation cumulative reward at each evaluation; the bottom panel shows the highest mean validation cumulative reward observed so far. Colours identify the same runs in both panels. In the bottom panel, dots mark the last improvement and crosses mark the final evaluation. Environment transitions are summed across the 6 parallel training environments.}
    \label{fig:validation_vs_environment}
\end{figure*}

We measure the mean cumulative reward for the validation set at intervals throughout the training. For the 9 models presented in \cref{sec:results}, \cref{fig:validation_vs_environment} plots the change in the mean cumulative reward for the validation set against the number of environment transitions. As discussed in \cref{sec:training and deployment}, training stops when the average cumulative reward for the validation set stagnates (does not improve for more than 5 measurements), or when $300,000$ environment transitions have occurred, whichever is sooner. We see that for all 9 seeds the training saturates before the hard limit of $300,000$. It is typical for the training to terminate after $\sim150,000$ transitions, with the lowest cumulative reward reached before $100,000$ transitions in the majority of cases.

\end{document}